\documentclass[11pt]{article}
\usepackage{amsmath,amssymb,color,graphics,epsfig,cite,tcolorbox}

\usepackage{amsfonts}

\newcommand{\hoch}[1]{$\, ^{#1}$}

\newcommand{\be}{\begin{equation}}
\newcommand{\ee}{\end{equation}}
\newcommand{\bea}{\setlength\arraycolsep{2pt} \begin{eqnarray}}
\newcommand{\eea}{\end{eqnarray}}
\newcommand{\nn}{\nonumber}

\def\ft#1#2{{\textstyle{\frac{\scriptstyle #1}{\scriptstyle #2} } }}
\def\fft#1#2{{\frac{#1}{#2}}}

\def\0{{\sst{(0)}}}
\def\1{{\sst{(1)}}}
\def\2{{\sst{(2)}}}
\def\3{{\sst{(3)}}}
\def\4{{\sst{(4)}}}
\def\5{{\sst{(5)}}}
\def\6{{\sst{(6)}}}
\def\7{{\sst{(7)}}}
\def\8{{\sst{(8)}}}
\def\sst#1{{\scriptscriptstyle #1}}

\begin{document}

\begin{center}
{\Large {\bf Regular Electric Black Holes from EMS Gravity}}

\vspace{20pt}

Zhi-Chao Li\hoch{1} and H. L\"{u}\hoch{1,2,3}

\vspace{10pt}

{\it \hoch{1}Center for Joint Quantum Studies and Department of Physics,\\
School of Science, Tianjin University, Tianjin 300350, China }

\bigskip

{\it \hoch{2}Joint School of National University of Singapore and Tianjin University,\\
International Campus of Tianjin University, Binhai New City, Fuzhou 350207, China}

\bigskip

{\it \hoch{3}Peng Huanwu Center for Fundamental Theory, Hefei, Anhui 230026, China}

\vspace{40pt}

\underline{ABSTRACT}
\end{center}

We construct Einstein-Maxwell-Scalar (EMS) theories that admit regular electric black holes. Such a Maxwell-scalar theory is equivalent to some nonlinear electrodynamics (NLED) at the level of equations of motion, but it has the advantage of circumventing the no-go theorem of regular electric black holes under a given Lagrangian of NLED. We study the thermodynamics and show that the mass of the regular black hole can be determined solely by the Maxwell field, without having to know the metric profile function. Our formalism allows to study the applications of the electrically-charged regular black holes in areas that were previously available only to the non-regular ones.

\vfill{lizc@tju.edu.cn\ \ \  mrhonglu@gmail.com}


\thispagestyle{empty}
\pagebreak

\tableofcontents
\addtocontents{toc}{\protect\setcounter{tocdepth}{2}}

\newpage
\section{Introduction}

Einstein's field equation relates the eponymous curvature tensor to matter energy-momentum tensor $T_{\mu\nu}$. If we do not impose conditions on $T_{\mu\nu}$, any spacetime could in principle arise. An important prediction of Einstein's theory is that black holes emerge within suitable $T_{\mu\nu}$. By contrast, the equally intriguing spacetime geometry, the traversable wormhole, such as the Ellis wormhole \cite{Ellis:1973yv}, requires exotic matter that violates the null energy condition (NEC). Under the strong energy condition (SEC), black hole geodesics were shown to be incomplete \cite{Hawking:1973uf, Senovilla:2014gza}; however, the nature of the incompleteness is not all clear. In most of explicit examples, including the Schwarzschild black hole, the singularity is in the form of the divergence of Riemann curvature tensor. There can also be impenetrable boundaries in the form of irrational power in the near-horizon geometry \cite{Mao:2023qxq}.

By relaxing the SEC, regular black holes with complete geodesics can be constructed. The first example was given by Bardeen \cite{2767662}, followed by Hayward \cite{Hayward:2005gi}. The geodesic completeness of the latter was under debate since the metric is not an even function of the radial coordinate $r$ and the geodesic can cross the de Sitter core and leak into the negative $r$ region where a curvature singularity resides \cite{Zhou:2022yio}. Both black holes satisfy the weak energy condition (WEC). Indeed, it was shown in \cite{Zaslavskii:2010qz} that the space-time structure of regular spherically-symmetrical black holes is linked to the violation of SEC, with the Tolman mass being negative in static regions under the event horizon, and in non-static regions, a different constraint leads to the violation of the dominant energy condition (DEC) for perfect fluids. However, spherically-symmetric regular black holes that satisfy the stronger DEC do exist. The first such example is perhaps the one constructed \cite{Cisterna:2020rkc} from the Einstein-quasi-topological electromagnetism \cite{Liu:2019rib}. Several classes of regular black holes satisfying DEC were obtained in \cite{Li:2023yyw}, where a new proof of the energy condition was given. All these regular black holes necessarily have de Sitter core. Regular black holes with Minkowski core were also constructed \cite{Xiang:2013sza}; however, it was shown in \cite{Li:2023yyw} that such regular black holes with special spherically-symmetric and static (SSS) ($g_{tt} g_{rr}=-1$) metrics necessarily violate the NEC.

One important aspect in regular black hole research is to construct a matter field theory as the source for the black hole, analogous to Einstein-Maxwell theory that admits the Reissner-Nordstr\"om (RN) black hole. Earlier examples of using suitable NLED theories to construct regular black holes were given in \cite{Ayon-Beato:1998hmi,Ayon-Beato:1999kuh,Ayon-Beato:1999qin}. In four dimensions, the Maxwell field can be sourced by either electric or magnetic charges, even when the theory becomes nonlinear. Regular black holes as magnetic monopoles in a suitable NLED are much easier to discuss. In particular, the NLED theory for the Bardeen black hole as a magnetic charged solution was obtained in \cite{Ayon-Beato:2000mjt}. In fact a reverse-engineer technique was developed \cite{Fan:2016hvf} for any SSS black holes, including regular ones. This technique allows one to construct magnetic charged black holes for a generic SSS metric. Their resulting theories, including the one for the Bardeen black hole, are unfortunately ugly in that they are typically nonanalytic and involve fractional powers of the square of the field strength, ${\cal F} = \fft14 F^{\mu\nu} F_{\mu\nu}$ with $F=dA$. In \cite{Li:2023yyw}, we constructed a few classes of analytic NLEDs that admit magnetic regular black holes. Furthermore, the theories at the weak field limit can be viewed as effective field theory (EFT) perturbations of Maxwell theory:
\be
L({\cal F}) = -{\cal F} +\alpha_1 {\cal F} ^2 + \alpha_2 {\cal F}^3 + \cdots\,.\label{wflimit}
\ee
Requiring $L({\cal F})$ to be an analytic function has an additional advantage that the metric function is guaranteed to even of $r$, hence ensuring the geodesic completeness at the core, both de Sitter and Minkowski. (Whether those magnetic ``regular'' black holes should be be viewed as regular solutions is controversial, since as pointed out in \cite{Li:2023yyw}, although the metric is regular, the quantity ${\cal F}$ diverges at the core.)

However, the success does not extend to the electrically-charged regular black holes for any given NLED $L({\cal F})$ theory. This is because for an electric solution, the quantity ${\cal F}$, which is negative, must vanish both at asymptotic infinity and at the core, implying there is at least one negative minimum ${\cal F}_{\rm min}$. In other words, ${\cal F}$ is not a monotonous function of radius $r$. Consequently, there must be multiple branches of the Lagrangian $L({\cal F})$ to describe such a regular black hole \cite{Bronnikov:2000vy}. Since Lagrangian is an off-shell concept, having multiple branches is clearly unacceptable in this formalism. No-go theorem was established in \cite{Bronnikov:2000vy, Bokulic:2022cyk}.

The electrically-charged regular black holes were thus constructed at the level of equations of motion, by introducing an ``auxiliary'' field $P_{\mu\nu}$ to carry the electric charge. Multiple branches emerge if one solves for $P$ in terms of ${\cal F}$, leading to multiple branches of $L({\cal F})$. In particular, the joining at the ${\cal F}_{\rm min}$ of the two branches forms an unattractive cusp \cite{Bronnikov:2000vy}.

In this paper, we resolve this predicament by considering Lagrangian formalism with an additional non-dynamic scalar field, in the form of
\be
{\cal L} = -\phi^{-1} {\cal F} - V(\phi)\,.\label{mslag}
\ee
The electromagnetic duality can be easily performed, yielding
\be
{\cal L} = -\phi\, {\cal G} - V(\phi)\,,
\ee
which was first proposed in \cite{Li:2023yyw} for magnetic solutions, mimicking the scalar-tensor theory of $f(R)$ gravity. In this approach, the fundamental fields are the scalar $\phi$ and Maxwell potential $A_\mu$, which are both single valued as functions of radius $r$ in the regular solutions. The multi-value arises if we want to express $\phi$ as a function ${\cal F}$, which is no longer a sensible option nor a requirement in the EMS theory.

In principle, the previous examples of the NLED theories (at the level of equations of motion) that admit electric regular black holes can now all have a proper Lagrangian of EMS in the form of \eqref{mslag}. But we find that these theories all have some shortcomings. Although the NLED for electric regular black holes cannot be written as a single Lagrangian of $L({\cal F})$, it can be done in the weak field perturbative region that determines the asymptotic infinity of the regular black holes. We find that in literature, many of NLEDs do not approach the Maxwell theory perturbatively. In the cases that they do, the sub-leading term involve fractional powers of ${\cal F}$, e.g.~\cite{Ayon-Beato:1998hmi, Ayon-Beato:1999kuh,Ayon-Beato:1999qin,Balart:2014cga}, which are poor examples as EFTs for the Maxwell theory.

In this paper, we would like to consider Einstein gravity coupled to Maxwell-scalar theories in the form of \eqref{mslag}, satisfying the following criteria
\begin{itemize}

\item The theories admit regular electric black holes.

\item Both the potential $V(\phi)$ and black hole metric can be explicitly written in terms of known special functions.

\item In the weak field perturbative region, the theories behave as \eqref{wflimit}.

\end{itemize}
The last criterium is motivated by the fact that there can be many interesting applications for electric black holes in both astrophysics and the AdS/CFT correspondence, and a sensible theory with no fractional power of ${\cal F}$ is more desirable. Furthermore, since Maxwell theory satisfy all the energy conditions, this approach leads theories that can satisfy the DEC or at least NEC more easily.

The paper is organized as follows. In section 2, we review the general properties of black holes charged under an NLED in the form of $L({\cal F})$. We show that the mass of the black hole can be independently determined by the Maxwell field, without having to know the complete metric solutions. In particular we obtain a formula for determine a critical value of mass so that the metric becomes regular or less singular. We also show that the theory can be cast into the form of Maxwell-scalar theory and thereby circumvent the no-go theorem of regular black holes charged under the NLED in the Lagrangian formalism. In section 3, we obtain new regular electric black holes in EMS theories from the previously known magnetic black holes by electromagnetic duality. We use a couple explicit examples to compare the common feature and difference between EMS and the corresponding NLED. In section 4, we construct directly several new classes of EMS theories that admit regular electric black holes and meet all the criteria discussed earlier. In section 5, we present a regular resolution to the RN black hole, where the Riemann tensor polynomials are smooth, but not their covariant derivatives. We call it an imperfect resolution. We conclude the paper in section 6. The EMS theories we construct are valid and general dimensions. In the appendix, we construct regular electric black holes of our EMS theories in higher dimensions.

Our work focuses narrowly on constructing a field theory for electric regular black holes. There have been tremendous progress in regular black hole research, including constructing solutions in pure Riemann tensor gravity \cite{Bueno:2024dgm}. We refer to the recent review \cite{Lan:2023cvz} for further references.

\section{From NLED to Maxwell-scalar theory}
\label{sec:NLED}

\subsection{Einstein-NLED gravity}

We begin with Einstein gravity minimally coupled to a special class of NLED described by the Lagrangian $L({\cal F})$, with the action
\be
I = \frac{1}{16 \pi} \int d^4 x \sqrt{-g}\big( R + L({\cal F})\big)\,,\qquad {\cal F}= \frac{1}{4}F_{\mu\nu}F^{\mu\nu}\,,\qquad F_{\mu\nu} = \partial_\mu A_\nu -\partial_\nu A_\mu\,.
\ee
The covariant equations of motion derived from varying the Maxwell field $A_\mu$ and the metric $g_{\mu\nu}$ are given by
\bea
\delta A_\mu:&&\qquad \nabla_\mu(L_{\cal F}F^{\mu\nu})=0\,,\qquad  L_{\cal F} =
\frac{\partial L}{\partial \mathcal{F}}\,; \nn\\
\delta g_{\mu\nu}:&&\qquad G_{\mu \nu} = \frac{1}{2}(L({\cal F})\, g _{\mu \nu} - L_{\cal F}\, F_{\mu\rho}F_{\nu}{}^{\rho})
\,,\label{covarianteq}
\eea
where $G_{\mu \nu} = R_{\mu \nu}-\frac{1}{2} R g_{\mu \nu} $ is the Einstein tensor.

The spherically-symmetric and static black hole solutions carrying electric and/or magnetic charges have the SSS metric; therefore, the most general ansatz in four dimensions is
\bea
ds^2 &=& -f(r) dt^2 + \frac{dr^2}{f(r)} + r^2(d\theta^2 + \sin^2 \theta d\phi^2)\,,\nn\\
A &=& -a(r) dt - p \cos \theta d \phi, \qquad dA=\psi\, dt \wedge dr + p \sin\theta d\theta \wedge d\phi, \qquad \psi=a'(r)\,.
\eea
Note that the SSS metric with $g_{tt} g_{rr}=-1$ is a consequence of the field equation $-G_{t}{}^{t}+G_{r}{}^{r}=0$, associated with our matter ansatz.

Having determined the ansatz, we are now in a position to solve the equations \eqref{covarianteq}.
The Maxwell equation is automatically satisfied for the magnetic ansatz, while the electric ansatz can be solved for $\psi=a'$, with an integration constant $q$. Specifically, we have
\be
\psi=\frac{q}{r^2 L_{\mathcal{F}}}\,,\qquad
\mathcal{F} = -\frac{1}{2} \psi^2 + \frac{p^2}{2 r^4}\,.\label{maxwellsol}
\ee
The integration constants $(q,p)$ parameterize the electric and magnetic charges, defined by
\be
Q_e= \frac{1}{16 \pi} \int L(\mathcal{F})\,{* F}=\fft14 q\,,\qquad
Q_m = \frac{1}{16 \pi} \int F = \fft14 p\,,\\\nn
\ee
It follows from the first equation in \eqref{maxwellsol} that the equation to determine the electric field $\psi$ is purely algebraic and independent of the metric function $f$. It is convenient to define
\be
{\cal F}_0= - \fft{q^2}{2r^4}\,,\qquad {\cal G}_0 = - \fft{p^2}{2r^4}\,,
\ee
which are the fiend strength squared value for the electric and magnetic RN black holes. It follows from the second equation of \eqref{maxwellsol}, the algebraic
equation becomes
\be
{\cal F}= \fft{{\cal F}_0}{L_{\cal F}} + {\cal G}_0\,.
\ee
For Maxwell theory, where $L_{\cal F}=1$, the solution is straightforward.
For NLED, the possibility of solving this equation analytically depends on the specific function \(L(\mathcal{F})\).

Having obtained the solution of the Maxwell equation for the NLED that is independent of the metric function, we can now solve the remaining Einstein equations. The three Einstein equations in the time, radial and 2-sphere directions are not independent, but they can all be solved provided that
\be
G_{t}{}^{t}+G_{r}{}^{r} = \frac{2(r f' + f -1)}{r^2} = L - 2 L_{\mathcal{F}} \mathcal{F}\,.\label{Einsteinsol}
\ee
The metric is thus given as an quadrature with an integration constant associated with the mass.

\subsection{Black hole thermodynamics}

\subsubsection{General properties}

The general solution described above generally contain three parameters, the mass $M$, the electric and magnetic charges $(Q_e,Q_m)$. For fixed charges and sufficiently large $M$, the solution describes a black hole with an event horizon $r_+$, which is the largest positive root of the metric profile $f(r)$. The temperature and entropy of the black hole are
\be
T= \fft{f'(r_+)}{4\pi}\,,\qquad S=\pi r_+^2\,.
\ee
These parameters satisfy the first law of black hole thermodynamics
\be
dM = T dS + \Phi_e dQ_e + \Phi_m dQ_m\,,\label{firstlaw}
\ee
where
\be
\Phi_e = a(r_+)=\int_{r_+}^\infty \psi(r) dr \,,\qquad \Phi_m = b(r_+) =
\int_{r_+}^\infty \fft{p}{r^2} L_{\cal F}\, dr\,.\label{phiephim}
\ee
Here, we introduced the magnetic potential $b(r)$, derived from the electromagnetic duality of NLED
\be
L_{\cal F}\, {* F}= G=d B\,,\qquad B=b(r) dt\,,
\ee
for spherically-symmetric ansantz. The magnetic component of $F$ give rise to electric $B$.

Note that the first law is guaranteed by the equations of motion via the Wald formalism \cite{Wald:1993nt}. Therefore, we can deduce some properties of the black hole without actually having to solve for them. This is particularly useful in NLED theory since not all of them will necessarily give analytic solutions, and yet some mass/charge relation can be determined without the explicit solutions. The first law \eqref{firstlaw} implies
\bea
\fft{\partial M}{\partial r_+} &=& 2\pi r_+ T = \fft12 r_+ f'(r_+)\,,\nn\\
\fft{\partial M}{\partial Q_e} &=& \Phi_e = a(r)\Big|^\infty_{r_+}=\int_{r_+}^\infty \psi(r) dr
= \int_{r_+}^\infty \frac{q}{r^2 L_{\mathcal{F}}} dr\,,\nn\\
\fft{\partial M}{\partial Q_m} &=& \Phi_m = b(r)\Big|_{r_+}^\infty=\int_{r_+}^\infty \frac{p\,L_{\mathcal{F}}}{r^2} dr\,.
\eea
These equations are significant since they shows that the mass of the black hole can be determined solely by the solution of the Maxwell field $\cal F$ of the NLED, which we have shown to be independent of the metric profile $f$. Since $f(r_+)=0$, $f'(r_+)$ can be derived from \eqref{Einsteinsol} locally without having to know the full solution. It is given by
\be
f'(r_+) = \frac{1}{r_+} + \frac{r_+}{2} (L - 2 L_{\mathcal{F}} \mathcal{F})\Big|_{r={r_+}}.\label{fprp}
\ee

\subsubsection{Example: EBI gravity}

For a concrete example, we consider the Einstein-Born-Infeld (EBI) gravity, where for purely electrically-charged black holes, the relevant part of BI theory \cite{Born:1934gh,Hoffmann:1935ty} is of the $L({\cal F})$ type, with
\be
L({\cal F}) = \fft{4}{\alpha} \Big(1-\sqrt{1 + \ft12 \alpha{\cal F}}\Big).
\ee
We have chosen the coupling constant such that it reduces to Maxwell theory in the $\alpha\rightarrow 0$ limit. For the purely electric ansatz, it follows from \eqref{maxwellsol} that we have
\be
\psi = \fft{q}{r^2\sqrt{1 + \fft{\alpha q^2}{4r^4}}}\,,\qquad \Phi_e = \int_{r_+}^\infty \psi = \frac{q}{r_+} \, _2F_1\left(\frac{1}{4},\frac{1}{2};\frac{5}{4};-\frac{\alpha q^2}{4 r_+^4}\right).\label{ebiphie}
\ee
It then follows from \eqref{fprp} that
\be
f'(r_+)=\frac{1}{r_+}-\frac{2 r_+}{\alpha } \left(\sqrt{1+\frac{\alpha  q^2}{4 r_+^4}}-1\right)\,.
\ee
We therefore have
\bea
\fft{\partial M}{\partial r_+} = \fft12 r_+ f'(r_+)\,,\qquad
\fft{\partial M}{\partial Q_e} = \Phi_e\,.
\eea
These two equations can be easily solved, giving
\be
M=\frac{r_+}{2}+\frac{r_+^3}{3 \alpha } \left(1-\, _2F_1\left(-\frac{3}{4},-\frac{1}{2};\frac{1}{4};-\frac{\alpha q^2}{4 r_+^4}\right)\right).
\ee
We thus obtain the mass as a function of the horizon radius $r_+$, without having to know the full analytic result of the metric profile $f$.

\subsection{Mass of regular or singularity-reduced black holes}

It was well known that for suitable NLED, one can construct regular black holes that has no curvature singularity from the origin $r=0$ to the asymptotic infinity. For small charges, smooth horizonless spacetime structures can also exist, for which the first law of black hole thermodynamics clearly no longer applies. Based on the Wald formalism, it was shown that an analogous first law of dynamics also exist, namely \cite{Liu:2020uaz,Li:2023yyw}
\be
dM= \Phi_e dQ_e + \Phi_m dQ_m\,, \label{fld}
\ee
where $\Phi_e$ and $\Phi_m$ are given by \eqref{phiephim} but with the lower limit of integration $r_+$ replaced by 0, namely
\be
\Phi_e =\int_{0}^\infty \psi(r) dr \,,\qquad \Phi_m =
\int_{0}^\infty \fft{p}{r^2} L_{\cal F}\, dr\,.\label{phiephimregular}
\ee
It was established \cite{Li:2023yyw} that the first law of dynamics \eqref{fld} holds even for regular black holes with event horizons. This therefore provides a powerful formula to calculate the mass of the regular black hole without having to know the metric function, since the electric or magnetic potentials can be derived independently of $f$:
\bea
\hbox{Magnetic:}\qquad M = \int_0^p \Big( \int_0^{\infty} -\frac{p\, L_{\mathcal{F}}}{r^2} dr \Big) dp\,,\nn\\
\hbox{Electric:}\qquad M = \int_0^q \Big( \int_0^{\infty} -\frac{q}{r^2 L_{\mathcal{F}}} dr \Big) dq\,.\label{regularmass}
\eea

This, however, leads to an interesting question. The electric or magnetic potential $\Phi_e$ or $\Phi_m$ can be regular at the origin $r=0$, even for singular black holes, such as the one in EBI gravity considered earlier. (As we can see from \eqref{ebiphie} that $\psi$ is integrable from $r=0$ to $\infty$.) Does the first law of dynamics \eqref{fld} still work? What is the physical implication of its mass derived from \eqref{fld}, or equivalently from \eqref{regularmass}? To understand this, we again use EBI gravity as an illustrative example. Replacing the lower limit of the integration in \eqref{ebiphie} with 0, we have
\be
\Phi_e=\frac{\sqrt{q}\, \Gamma \left(\frac{1}{4}\right)^2}{2 \sqrt{2 \pi } \sqrt[4]{\alpha }}\,,\qquad M=M_{\rm cr}\equiv \int \Phi_e dQ_e=\frac{q^{3/2} \Gamma \left(\frac{1}{4}\right)^2}{12 \sqrt{2 \pi } \sqrt[4]{\alpha }}\,.\label{ebimasscr}
\ee
To understand what happens to the charged black hole when the mass takes this critical value $M_{\rm cr}$, we examine the metric function $f$ for the charged black hole, whose analytic solution exists, given by
\be
f(r)= 1-\frac{2 M}{r}-\frac{2 r^2}{3 \alpha}\Big(1-\, _2F_1\Big(-\frac{3}{4},-\frac{1}{2};\frac{1}{4};-\frac{\alpha q^2}{4 r^4}\Big)\Big)\,.
\ee
The power-series expansions at $r=0$ and at large $r$ are
\bea
r\rightarrow 0:&&\qquad f \sim - \fft{2(M-M_{\rm cr})}{r} + 1
-\frac{q}{\sqrt{\alpha }}+\frac{2 r^2}{3 \alpha }+\cdots\,,\nn\\
r\rightarrow \infty:&& \qquad f \sim 1 - \fft{2M}{r} + \fft{q^2}{4r^2} -\frac{\alpha q^4}{320 r^6} + \cdots
\eea
Thus we see that when $M=M_{\rm cr}$, the divergent term at $r=0$ in the metric function vanishes, giving rise to a black hole that is less singular. This critical value of mass and its related properties were known to \cite{Li:2016nll}, but it was not known that it could be obtained from \eqref{ebimasscr}.

In general, if the electric field $\psi$ is regular at $r=0$ in some NLED theory, its contribution to equation of the metric profile $f$ \eqref{Einsteinsol} may not be singular. Therefore, we should have the following general power-series expansion at $r=0$:
\be
f=-\fft{2(M-M_{\rm cr})}{r} + 1 + c_0 + c_1 r + c_2 r^2 + c_3 r^3 + c_4 r^4 + \cdots\,,\qquad r\rightarrow 0
\ee
(Here, we assume that the NLED theory is sufficiently well-behaved; we shall come back to this point later.)  The Riemann tensor squared behaves as
\bea
\hbox{Riem}^2 &=& \frac{1}{r^4}\Big(r^4 f''(r)^2+4 r^2 f'(r)^2+4 f(r)^2-8 f(r)+4\Big)\nn\\
&=&\frac{48 \left(M-M_{\rm cr}\right){}^2}{r^6}-\frac{16 c_0 \left(M-M_{\rm cr}\right)}{r^5}+\frac{4 c_0^2}{r^4}+\frac{8 c_0 c_1}{r^3}\cr
&&+\frac{8 \left(c_1^2+c_0 c_2-2 c_3 \left(M-M_{\rm cr}\right)\right)}{r^2}+\frac{8 \left(3 c_1 c_2+c_0 c_3-6 c_4 \left(M_-M_{\rm cr}\right)\right)}{r}+O\left(1\right).
\eea
Thus, we see that taking the critical value of mass reduces the order of power-law singularity from 6 to 4. This is exactly what happens in the case of charged black hole in EBI gravity.  In order to construct regular black holes, we see that in addition to setting $M=M_{\rm cr}$, the function $f$ should have $c_0=0=c_1$. For the EBI charged black hole, we have $c_0=-q/\sqrt{\alpha}$ and $c_1=0$, yielding a reduced $1/r^4$ singularity from the general $1/r^6$ singularity. Of course, in the case of Einstein-Maxwell gravity, where $\psi=q/r^2$ is singular at $r=0$, the above argument breaks down, and the electric field contributes a $1/r^8$ singularity at $r=0$ for Riem$^2$. Finally, it is important to emphasize that the above discussion is independent of whether the metric has an event horizon or not.

\subsection{The Maxwell-scalar theory}

As was discussed in the introduction, the NLED theory that admits the electrically charged regular black holes does not have a Lagrangian formalism, since it would involve multiple branches, which contradict its off-shell nature.

One way to resolve this issue is to observe that an NLED $L({\cal F})$ is equivalently to a scalar-vector theory \cite{Li:2023yyw}, analogous to the relation between $f(R)$ gravity and the scalar-tensor theory. We can thus adopt the $f(R)$ technique and write the equivalent Maxwell-scalar theory
\begin{equation}
L = -\phi\, \mathcal{F} - V(\phi)\,,\label{svt}
\end{equation}
where $\phi = -L_\mathcal{F}$ and $V(\phi) = L-L_\mathcal{F} \mathcal{F}$. There is a correspondence mapping of $L\leftrightarrow V$ and ${\cal F}\leftrightarrow \phi$ in that
\be
\fft{\partial L}{\partial {\cal F}} = - \phi\,,\qquad \fft{\partial V}{\partial \phi}=-{\cal F}\,.\label{correspondence}
\ee
For a given NLED $L({\cal F})$, we can use the left equation in \eqref{correspondence} to solve ${\cal F}$ in terms of $\phi$ and substitute it into the right equation and integrate $V$ as a function $\phi$. On the other hand, if we are given a speciific $V(\phi)$, we can use the right equation to determine $\phi$ in terms of ${\cal F}$, and substitute it into the left equation to integrate $L$ in terms of ${\cal F}$ and obtain the $L({\cal F})$ theory.

In this Maxwell-scalar description, the multiple $L({\cal F})$'s now belong to different branches of solutions to the algebraic equation of $\phi$ (the right equation in \eqref{correspondence}) in the same Lagrangian. This resolves the issue of having to use multiple Lagrangians to describe one solution. We only have one Lagrangian in the scalar-vector description. Another advantage of this scalar-vector theory is that it becomes straightforward to perform electromagnetic duality transformations since the theory becomes linear in the Maxwell field strength. The dual theory of \eqref{svt} is simply given by
\begin{equation}
    L_{\rm dual} = -\phi^{-1}\, \mathcal{F} - V(\phi)\,.\label{svtdual}
\end{equation}

\section{Electric regular black holes from electromagnetic duality}
\label{sec:emduality}

\subsection{General discussion}

In our previous paper \cite{Li:2023yyw}, we have constructed three classes of NLED theories in the form of $L({\cal G})$ that admit magnetic regular black holes. They are:
\begin{equation}
    L_1= -\frac{{\cal G}}{(1 + \left(\alpha {\cal G}\right)^n)^{\frac{1}{n}}},\quad
    L_2 = \frac{1}{\nu \alpha}  \left(\frac{1}{\left(1 + \alpha {\cal G}\right)^\nu }-1\right), \quad
    L_3 = \frac{1}{2\alpha}  \left(e^{-2\alpha {\cal G}} - 1\right). \label{variousL}
\end{equation}
Note the notational change here. Since these theories are intended to construct magnetic solutions, we use $G=dB$ for the Maxwell field strength and define ${\cal G}=\fft14 G^{\mu\nu} G_{\mu\nu}$. (Do not confuse this $G_{\mu\nu}$ with the Einstein tensor.) These theories are all analytic and can satisfy the DEC for appropriate parameters, with weak field limit as \eqref{wflimit}.

These theories all admit magnetically charged regular black holes. One could in principle obtain the regular electric solutions by performing electromagnetic duality. However, a direct discussion of such a duality can be complicated for these NLED's. Instead, it was observed in \cite{Li:2023yyw} that
these three classes of theories can all be cast into the Maxwell-scalar theory, namely
\begin{equation}
    L({\cal G}, \phi) = -\phi\, \mathcal{G} - V(\phi),
\end{equation}
with the corresponding scalar potentials respectively given by \cite{Li:2023yyw}
\begin{eqnarray}
V_1 &=&  \frac{1}{\alpha} \left(1 - \phi^\frac{n}{n+1} \right)^{\frac{n+1}{n}},\qquad n=1,2,\ldots,\nonumber\\
V_2 &=& \frac{1}{\alpha \nu} \left(1 + \nu \phi-(\nu + 1) \phi^\frac{ \nu}{\nu + 1}\right),\nonumber\\
V_3 &=& \frac{1}{2\alpha} \left(1 - \phi + \phi \log \phi\right). \label{variousphi}
\end{eqnarray}
Now it is straightforward to perform the electromagnetic duality and the resulting Maxwell-vector theories take the form
\begin{equation}
    L({\cal F}, \phi) = -\phi^{-1}\mathcal{F} - V(\phi)\,.\label{escalarvector}
\end{equation}
The new theories then admit regular electric black holes as the electromagnetic duals of the corresponding magnetic ones.

\subsection{Electrically-charged regular black holes}

Having obtained the Maxwell-scalar theory \eqref{escalarvector} with three classes of scalar potentials \eqref{variousphi}, it is straightforward to construct spherically-symmetric and static electrically-charged black holes, with the ansatz
\be
ds^2 = - f(r) dt^2 + \fft{dr^2}{f(r)} + r^2 (d\theta^2 + \sin^2\theta d\phi^2)\,,\quad
A=a(r) dt + p \cos\theta\,d\phi\,,\quad \phi=\phi(r)\,.
\ee
The scalar and Maxwell equations imply
\bea
\hbox{Scalar:}&&\qquad V'(\phi)=\phi^{-2} {\cal F}\,,\label{scalareq}\\
\hbox{Maxwell:}&&\qquad a'\equiv\psi= \fft{q\phi}{r^2}\,,\quad\rightarrow\quad {\cal F} = \phi^2 {\cal F}_0 + \fft{p^2}{2r^4}\,,\label{dyoncf}
\eea
where we define
\be
{\cal F}_0 \equiv -\fft{q^2}{2r^4}\,.\label{cf0def}
\ee
Note that ${\cal F}_0$ is precisely the corresponding fiend strength squared in the RN black hole; therefore, it provides us a fiducial point as how we should deviate from the Maxwell theory.

These two equations allow us to completely determine both $(\phi, {\cal F})$ in terms of the radial coordinate $r$, via ${\cal F}_0$. The metric profile function can be determined by the first-order differential equation
\be
\frac{r f' + f -1}{r^2} = -\fft12 V - \fft1{4r^4} (\phi q^2 + \phi^{-1} p^2)\,.\label{eqf}
\ee
For our current purpose, we set the magnetic charge parameter $p=0$.

Following the above steps, we can easily obtain the full electric solutions for the three classes of potentials in \eqref{variousphi}. Specifically, we have
\bea
V_1:\qquad \phi &=& \frac{1}{(1+(-\alpha \mathcal{F}_0)^{n})^{1+\frac{1}{n}}}\,,\qquad {\cal F}=\fft{{\cal F}_0}{(1+(-\alpha \mathcal{F}_0)^{n})^{2+\frac{2}{n}}}\,,\nn\\
f &=& 1 - \frac{2 M}{r}+\frac{q^2 }{4 r^2} \, _2F_1 (\frac{1}{4n},\frac{1}{n};1+\frac{1}{4n};(-\frac{\alpha q^2}{2 r^4})^n ) \,;\label{v1sol}\\
V_2:\qquad
\phi &=& \frac{1}{(1-\alpha \mathcal{F}_0)^{1+\nu}}\,,\qquad {\cal F}= \fft{{\cal F}_0}{(1 - \alpha {\cal F}_0)^{2+2\nu}}\,,\nn\\
f &=& 1 - \frac{2 M}{r}+\frac{r^2 }{6  \alpha \nu}\big(\, _2F_1 (-\frac{3}{4},\nu;\frac{1}{4};-\frac{\alpha q^2}{2 r^4} )-1\big)\,;
\label{v2sol}\\
V_3:\qquad \phi&=& e^{2 \alpha \mathcal{F}_0 }\,,\qquad {\cal F}={\mathcal{F}}_0 e^{4 \alpha \mathcal{F}_0 }\,,\nn\\
f &=& 1 - \frac{2 M}{r}-\frac{r^2}{12 \alpha} + \frac{q^{3/2} }{16 \alpha^{1/4} r} \big(\Gamma (-\frac{3}{4},\frac{\alpha q^2}{r^4} )- \Gamma (-\frac{3}{4})\big)\,.\label{v3sol}
\eea
Note that in the first line, $n$ is an integer, and in all these examples, $\phi$ and $\cal F$ are analytic functions of the radial coordinate $r$. These black holes all have two independent parameters, namely the mass $M$ and the electric charge $Q_e=q/4$; therefore, they are generally singular at $r=0$. It follows from the expressions of ${\cal F}$ that the electric field strength $\psi=\sqrt{-2{\cal F}}$ is regular from $r=0$ to asymptotic infinity. As we have discussed in section \ref{sec:NLED}, in this case, there exists a critical $M_{\rm cr}$, defined by
\be
M= \int_{0}^Q dQ \Big(\int_0^\infty \psi dr\Big) =
\int_{0}^q dq \Big(\int_0^\infty \fft14 \sqrt{-2\cal F} dr\Big) \,.\label{MM}
\ee
For the three classes of solutions, we find
\bea
V_1:\qquad M_{\rm cr}&=&\frac{\sqrt[4]{2} q^{3/2} \Gamma \left(1+\frac{1}{4 n}\right) \Gamma \left(1+\frac{3}{4 n}\right)}{6 \sqrt[4]{\alpha} \Gamma \left(1+\frac{1}{n}\right)}\,,\nn\\
V_2: \qquad M_{\rm cr}&=& \frac{q^{3/2} \Gamma \left(\frac{1}{4}\right) \Gamma \left(\nu+\frac{3}{4}\right)}{12\ 2^{3/4} \sqrt[4]{\alpha} \nu \Gamma (\nu)}\,,\nn\\
V_3: \qquad M_{\rm cr}&=& \frac{q^{3/2} \Gamma \left(\frac{1}{4}\right)}{24 \sqrt[4]{\alpha}}\,.
\eea
It can be easily verified that for all the three classes discussed above, the metric profile function $f$ at $r=0$ behaves as
\be
f=1 + \fft{M-M_{\rm cr}}{r} + {\cal O}(r^2)\,.
\ee
Therefore, these electrically charged black holes are all regular at $r=0$, when the mass is precisely given by the corresponding critical value. This again illustrates that the mass of these regular black holes can be determined without having to solve for $f$. Note that for $V_1$ and $V_2$ examples, the regular black holes have de Sitter core and can satisfy the DEC. For $V_3$, they have instead the Minkowski core, and hence all violate the NEC.

\subsection{As effective NLED $L({\cal F})$}

The scalar equation \eqref{scalareq} is algebraic for $\phi$, which can be solved in terms of ${\cal F}$, and it is consistent to substitute the solution back into the Lagrangian \eqref{escalarvector} and obtain an NLED in the form of $L({\cal F})$. In the literature, however, there are no-go theorems on regular electric black holes from NLED of the type $L({\cal F})$ \cite{Bronnikov:2000vy, Bokulic:2022cyk}. It is thus instructive to investigate the difference between our Maxwell-scalar theory and the NLED. In this subsection, we study the resulting NLED and illustrate its pathology. For the three classes of $V(\phi)$ given in \eqref{variousphi}, the scalar equation \eqref{scalareq} can only be solved analytically for very limited cases. We find two relatively simple explicit examples, and we shall present them below to illustrate the pathology of $L({\cal F})$ in constructing electric regular black holes.

\bigskip
\noindent{\bf Example 1}
\medskip

The first example is provided by $V_2$ in \eqref{variousphi} with $\nu=1/2$, namely
\be
V=\fft{1}{\alpha} \Big(2  + \phi - 3 \phi^{\fft13}\Big)\,.\label{scalarpot1}
\ee
The black hole solutions were given in \eqref{v2sol} with $\nu=1/2$.
For the scalar potential \eqref{scalarpot1}, the scalar equation \eqref{scalareq} is essentially a cubic polynomial equation of $\phi^{2/3}$:
\be
(\phi^\fft23)^3-(\phi^{\fft23})^2 -\alpha  {\cal F} =0\,.
\ee
It can be solved exactly, and we find that two branches of the solutions are relevant, given by
\be
\phi_k ({\cal F}) =\Big[\fft13+ \fft23 \cos\Big(\frac{1}{3} \arccos(1 + \frac{27}{2} \alpha
{\cal F}) - \frac{2}{3}\pi k\Big)\Big]^{\fft32},\qquad
k=0, 1\,.
\ee
Substituting these two branches of $\phi$ leads to two branches of NLED, namely
\be
L_k({\cal F}) = -\phi_k^{-1} {\cal F} - \fft{1}{\alpha} \Big(\phi_k - 3 \phi_k^{\fft13} +2\Big),\qquad k=0,1.
\ee
For our case where $F_{\mu\nu}$ is electric, the allowed range of ${\cal F}$ is $-\fft{4}{27\alpha} \equiv {\cal F}_{\rm min} \le {\cal F}\le 0 $, with the minimum occuring at $r=r_*$, with
\be
r_*^2=\sqrt{\alpha}\,q\,.
\ee
The inner region $0\le r< r_*$ is described by the $k=1$ branch, while the outer region $r_*<r< \infty$ is described by the $k=0$ branch. In both cases, the function ${\cal F}$ lies in the region $( {\cal F}_{\rm min},0)$. This is certainly unaccepted for the path integration formulation of a quantum field theory, since the formalism is based on the off-shell principle. However, this problem can be solved by introducing the auxiliary scalar $\phi$, where the multiple branches arise only in the classical on-shell situations. Nevertheless, it can still be useful  to study the effective theory on its own for either the inner or outer regions.

Note that for small ${\cal F}$, we have
\bea
k=0:&& L=-\mathcal{F}+\frac{3 \alpha  \mathcal{F}^2}{4}-\frac{13 \alpha ^2 \mathcal{F}^3}{8}+O\left(\mathcal{F}^4\right), \cr
k=1:&& L=-\fft{2}{\alpha} \Big(1  -2 (-\alpha \mathcal{F})^{\fft14} +\frac{1}{2} (-\alpha \mathcal{F})^{\fft34}-\frac{3}{16}(-\alpha \mathcal{F})^{5/4}+ O(\mathcal{F}^{3/2})\Big).
\eea
Thus the branch describing the outer region can be viewed as a perturbative effective field theory of the Maxwell theory, with only integer powers of ${\cal F}$. In the inner region, however, the theory is highly non-perturbative.
It is also worth pointing out that $(\phi_1,\phi_2)$ are continuous at the minimum  ${\cal F}_{\rm min}$, but their derivatives diverge. Consequently $
L_1$ and $L_2$, as functions of ${\cal F}$ are continuous at ${\cal F}_{\rm min}$, but not their derivatives. The behaviors of the inner and outer $L({\cal F})$ was roughly plotted in the left panel of Fig.~\ref{figures}; they form a cusp at the joining. These types of properties were first pointed in \cite{Bronnikov:2000vy} while analysing the regular electric black holes of \cite{Ayon-Beato:1998hmi,Ayon-Beato:1999kuh,Ayon-Beato:1999qin}.

\bigskip
\noindent{\bf Example 2}
\medskip

The second example is provided by the scalar potential $V_3$ in \eqref{variousphi}. The scalar equation \eqref{scalareq} is given by
\be
\phi ^2 \log \left(\phi ^2\right)=4 \alpha  {\cal F}\,.
\ee
The solutions are given by product-log functions that can have multiple branches. We find that there need to be two branches, given by
\begin{equation}
\phi_k({\cal F})  = e^{\fft12 W_k(4\alpha {\cal F})}\quad\rightarrow\quad
L_k=-\frac{2}{\alpha}\big(4\alpha {\cal F} e^{-\fft12 W_k(4\alpha {\cal F})}- e^{\fft12 W_k(4\alpha {\cal F})}+1\big)\,,
\end{equation}
with $k=0,-1$. For small $\alpha {\cal F}$, the two branches behave as
\bea
k=0:&& L=-\mathcal{F}+\frac{ \alpha  \mathcal{F}^2}{4}-\frac{5 \alpha ^2 \mathcal{F}^3}{24}+O\left(\mathcal{F}^4\right), \cr
k=1:&& L=-\frac{2}{\alpha}+\frac{2}{\alpha} (\alpha \mathcal{F})^{\ft12}+O\big(\mathcal{F}^{\ft32}\big).
\eea
Thus, the $k=0$ branch describes the perturbative outer branch, and the $k=1$ branch describes the non-perturbative inner branch. The two effective Lagrangians $L_k({\cal F})$ are continuous at the joint $\alpha {\cal F}=-1/(4 e)$, but their derivatives are divergent at the joining with opposite signs. The behavior was roughly plotted in the left panel of Fig.~\ref{figures}.

It should be remarked that although multi-branches of ${\cal L}({\cal F})$ imply the pathology of the Lagrangian formalism of NLED, it can still be useful if we want to focus the study on the isolated region.

\section{Constructing regular electric black holes directly}
\subsection{The setup}

We have obtained three classes of EMS theories that admit electrically-charged regular black holes by performing the electromagnetic duality on the previously-known magnetic solutions. In fact, the formalism we presented allowed us the construct the theories on its own, without having to rely on the regular magnetic construction. It follows from the earlier discussion that both the scalar potential $V(\phi)$ and metric profile $f$ are determined once the electric field, or ${\cal F}$, is given. Thus, the construction of the EMS theory that allows a regular electric black hole amounts to a sensible ansatz for ${\cal F}$.  The choice of ${\cal F}={\cal F}_0$, defined by \eqref{cf0def}, leads to the RN black hole of Einstein-Maxwell gravity. We would want the black hole to approach the RN black hole asymptotically. This leads to a general ansatz for ${\cal F}$:
\be
{\cal F}\quad \rightarrow\quad \left\{
    \begin{aligned}
  &\phantom{Xx} {\cal F}_0\,, \quad\qquad      r\rightarrow \infty\,; \\
  &\fft{{\cal F}_0}{(-\alpha {\cal F}_0)^{\nu}}\,,  \qquad   r\rightarrow 0\,,
    \end{aligned}
    \right.\label{cFsmallr}
\ee
where the parameter $\nu$ is to be determined so that the black hole is free from curvature singularity at $r=0$. Since ${\cal F}_0$ diverges at $r=0$, it is natural to expect that regularity at the origin would require $\nu$ to be positive and bigger than 1 so that ${\cal F}$ is finite at $r=0$. To determine the precise lower bound of $\nu$, we use the previously established procedure. We find that, for the ${\cal F}$ given in the second line above, the corresponding scalar potential is
\be
V \quad =\quad
\left\{
    \begin{aligned}
  &- \fft{\nu}{(\nu-2)\alpha} \phi^{\fft{\nu-2}{\nu}}\,, \qquad\quad \nu\ne 2\,; \\
  & - \fft{1}{\alpha} \log(\phi)\,,\qquad\qquad\quad \nu=2\,.
    \end{aligned}
    \right.
\ee
The corresponding metric profile function is
\be
f = \left\{
    \begin{aligned}
       & 1-\frac{2 m}{r} -\frac{r^2{\cal F}_0 \left(-\alpha  {\cal F}_0\right)^{-\frac{\nu}{2}}}{(\nu-2) (2 \nu-1)}\qquad\qquad \nu\ne 2\,;\\
&1-\frac{2 m}{r}-\frac{r^2(7+3\log \left(-\alpha  {\cal F}_0\right))}{18 \alpha }\qquad\qquad \nu=2\,.
    \end{aligned}
\right.
\ee
Thus we see that as $r\rightarrow 0$, the metric is regular for $\nu>2$, with the lower bound $\nu=2$ corresponding to a marginal logarithmic divergence.

As we discussed in Introduction, we further require that the effective NLED at the asymptotic region should behave as \eqref{wflimit}.  This implies that ${\cal F}$ should follow a Taylor series expansion of ${\cal F}_0$ in the asymptotic large-$r$ region, namely
\be
{\cal F} = {\cal F}_0 \Big(1+ c_i \sum_{i=1} (-\alpha {\cal F}_0)^i\Big)\,,\qquad \hbox{for}\qquad r\rightarrow \infty\,.\label{cFlarger}
\ee
We therefore have obtained the condition on desired ${\cal F}({\cal F}_0)$ for both the $r\rightarrow 0$ region, which ensures regularity, and the $r\rightarrow \infty$ region, which ensures that the solution is an analytic perturbation to the RN black hole.

We thus have a well-defined prescription to construct Maxwell-scalar theory that admits electrically charged regular black holes. First, we consider a suitable ansatz for ${\cal F}={\cal F}({\cal F}_0)$, from which we obtain
\be
\phi^2= \fft{{\cal F}}{{\cal F}_0}\,.
\ee
This also means that ${\cal F}_0$ is a function of $\phi$, namely ${\cal F}_0=
{\cal F}_0(\phi)$. It then follows from \eqref{scalareq} that the scalar potential can be determined by the simple first-order equation
\be
V'(\phi)= \phi^{-2} {\cal F} = {\cal F}_0(\phi)\,.\label{Vprimephi}
\ee
The scalar potential can then be solved. Note that our procedure allows to introduce the cosmological constant $\Lambda$ in a straightforward way, since the above equation is invariant under $V\rightarrow V+ 2\Lambda$.

\subsection{Examples}

We now consider explicit examples. The simplest ${\cal F}_0$ that satisfies \eqref{cFsmallr} and \eqref{cFlarger} is perhaps
\be
{\cal F}= \frac{{\cal F}_0}{(1 - \alpha {\cal F}_0)^{2\nu+2}}\,,\qquad \nu > 0\,.
\ee
This case was already presented in section 3, but worth presenting as an simple exercise. Following the general procedure discussed earlier, it is easy to show that the corresponding scalar potential is indeed given by the second case in \eqref{variousphi}. A new example is provided by
\be
{\cal F}= \frac{{\cal F}_0}{1 + (-\alpha {\cal F}_0)^n}\,,\qquad n=3, 4, 5, \ldots\,.
\ee
The scalar potential now is more complicated, expressed in terms of the incomplete Euler-$\beta$ function:
\be
V=-\fft{1}{2\alpha}\left(B_{\phi ^2}\left(\frac{1}{2}-\frac{1}{n},1+\frac{1}{n}\right)+\frac{2 \Gamma \left(\frac{1}{2}-\frac{1}{n}\right) \Gamma \left(1+\frac{1}{n}\right)}{\sqrt{\pi }}\right).
\ee
The scalar field of the charged black hole is $\phi = \sqrt{{\cal F}/{\cal F}_0}$, and the metric function $f$ is
\bea
f &=& 1 - \fft{2M}{r} \nn\\
&&+ \fft{q^2}{12 r^2} \Big(4 \, _2F_1\Big(\frac{1}{2},\frac{1}{4 n};1+\frac{1}{4 n};-(-\alpha {\cal F}_0)^n\Big)-\, _2F_1\Big(\frac{1}{2},\frac{1}{n};1+\frac{1}{n};-(-\alpha{\cal F}_0 )^n\Big)\Big).
\eea
It is easy to verify that the critical mass for regularity at $r=0$ is given by
\be
M_{\rm cr}=
\int_{0}^q dq \Big(\int_0^\infty \fft14 \sqrt{-2\cal F} dr\Big) =\frac{q^{3/2} \Gamma \left(\frac{1}{2}-\frac{1}{4 n}\right) \Gamma \left(1+\frac{1}{4 n}\right)}{3\ 2^{3/4} \sqrt{\pi } \sqrt[4]{\alpha }}\,.
\ee

The third example groups the above two cases together and we consider
\be
{\cal F}= \frac{{\cal F}_0}{(1 + (-\alpha {\cal F}_0)^n)^{\nu}}\,,\qquad n \nu >2\,,
\label{nnuclass}
\ee
with $n$ being positive integers. The corresponding scalar potential is
\be
V=\frac{\nu}{2 \alpha} \left(B_1\left(\frac{\nu }{2}-\frac{1}{n},1+\frac{1}{n}\right)-B_{\phi ^{2/\nu }}\left(\frac{\nu }{2}-\frac{1}{n},1+\frac{1}{n}\right)\right)\label{nnuV}
\ee
The metric function $f$ is
\bea
f&=&1-\frac{2M}{r}\nn\\
&&-\frac{q^2 }{12 r^2}\left(\, _2F_1\left(\frac{1}{n},\frac{\nu}{2};1+\frac{1}{n};-(-\alpha {\cal F}_0)^n\right)-4 \, _2F_1\left(\frac{1}{4 n},\frac{\nu}{2};1+\frac{1}{4 n};-(-\alpha {\cal F}_0)^n\right)\right).
\eea
The critical mass for the regular black hole is given by
\be
M_{\rm cr}=
\int_{0}^q dq \Big(\int_0^\infty \fft14 \sqrt{-2\cal F} dr\Big) =\frac{q^{3/2} \Gamma \left(1+\frac{1}{4 n}\right) \Gamma \left(\frac{2 n \nu-1}{4 n}\right)}{3\ 2^{3/4} \sqrt[4]{\alpha} \Gamma \left(\frac{\nu}{2}\right)}.
\ee
Note that $\nu=2 + 1/n$ case was already obtained via electromagnetic duality in section \ref{sec:emduality}.

The final explicit example we consider is a different class, where ${\cal F}$ is an exponential function of ${\cal F}_0$:
\be
{\cal F}= {\cal F}_0 e^{-\frac{(-\alpha \mathcal{F}_0)^n}{n}}\,,
\label{eclass}
\ee
where $n$ are positive integers. The scalar potential is then a gamma function of $\log\phi$:
\be
V=\frac{2^{1/n}}{\alpha } \left(\Gamma \left(1+\frac{1}{n}\right)-\Gamma \left(1+\frac{1}{n},-\log (\phi )\right)\right).
\label{ve}
\ee
The metric function is
\bea
f &=& 1 -
 \Big(2 M-\frac{2^{\frac{1}{4} \left(\frac{1}{n}-7\right)} q^{3/2} \Gamma \left(\frac{1}{4 n}\right)}{3 \sqrt[4]{\alpha } n}\Big)\fft1{r}\nn\\
&& -\frac{2^{\frac{1}{n}-1} r^2 \left(\Gamma \left(\frac{1}{n}\right)-\Gamma \left(\frac{1}{n},\fft12 (-\alpha {\cal F}_0)^n\right)\right)}{3 \alpha  n}
-\frac{q^2 E_{1-\frac{1}{4 n}}\left(\frac{1}{2}(-\alpha {\cal F}_0)^n\right)}{12 n r^2}\,.
\eea
The critical mass for regular black hole is
\be
M_{\rm cr}=
\int_{0}^q dq \Big(\int_0^\infty \fft14 \sqrt{-2\cal F} dr\Big) =\frac{2^{\frac{1}{4} \left(\frac{1}{n}-11\right)} q^{\fft32} \Gamma \left(\frac{1}{4 n}\right)}{3n \sqrt[4]{\alpha}}\,.
\ee
Note that the $n=1$ case was obtained earlier via electromagnetic duality in section \ref{sec:emduality}.

In all these examples, the scalar potentials have a fixed point at $\phi=1$. We have chosen the integration constant of \eqref{Vprimephi} so that $V(\phi=1)=0$. The cosmological constant, as an integration constant of \eqref{Vprimephi} can be easily introduced, which we discuss later.

\subsection{Marginal logarithmic singularity}

In the general class of \eqref{nnuclass}, regularity requires that $n \nu>2$. We now examine the solution when the bound is saturated, namely $n \nu=2$. The metric function is now given by
\bea
f &=& 1- \fft{2M}{r} -\frac{q^2 \big(1+(-\alpha  {\cal F}_0)^n\big)^{-1/n}}{12 r^2} +\frac{q^2}{12 r^2}\bigg(4 \, _2F_1\Big(\frac{1}{4 n},\frac{1}{n};1+\frac{1}{4 n};-(-\alpha {\cal F}_0 )^n\Big)\nn\\
&&-\frac{(-\alpha {\cal F}_0)^n}{n+1}\, _2F_1\Big(1+\frac{1}{n},1+\frac{1}{n};2+\frac{1}{n};-(-\alpha {\cal F}_0)^n\Big)\bigg).
\eea
The critical mass is given by
\be
M_{\rm cr}=\frac{q^{3/2} \Gamma \left(1+\frac{1}{4 n}\right) \Gamma \left(\frac{3}{4 n}\right)}{3\ 2^{3/4} \sqrt[4]{\alpha } \Gamma \left(\frac{1}{n}\right)}\,.
\ee
As $r$ approaches zero, we find that
\be
f= \fft{2(M-M_{\rm cr})}{r} + 1 - \frac{3 \log (-\alpha  {\cal F}_0)+7-\frac{3}{n} H_{\frac{1}{n}}}{18 \alpha }\, r^2 + {\cal O}(r^4)\,.
\ee
Owing to the logarithmic term at the $r^2$ order, the curvature at $r=0$ is divergent even when we take $M=M_{\rm cr}$. The divergence however is mild, in that for the $n$'th-order Riemann tensor polynomial, the leading divergence is $(\log r)^n$.

\subsection{More useful formulae}

In all the explicit examples we considered in this section, the metric function $f$ involves hypergeometric functions or other complicated special functions. It is formally obtained from the integration of the first-order equation \eqref{eqf}. However, in these examples, the source on the right-hand side of the equation can already be complicated, involving hypergeometric functions. One has to be very fluent in special functions to know that this first-order equation can be explicitly integrated. Here, we provide further useful formulae so that we can solve the systems of equation in different routes.

It is instructive to introduce quasi-local mass by $f=1-2m(r)/r$. It is easy to show that $m(r)$ satisfies
\be
m'(r) = -\fft{r^2}{4} \Big(L + 2\sqrt{{\cal F} {\cal F}_0}\Big)\,.\label{mpr}
\ee
The second term on the right-hand side of the equation is completely determined by the ansatz in our construction. The quantity $L$ is the on-shell Lagrangian, satisfying
\be
\frac{d L}{d r} = -\phi^{-1} \frac{d {\cal F}}{d r} = - \sqrt{\frac{{\cal F}_0}{{\cal F}}} \frac{d {\cal F}}{d r}\,.
\ee
Thus, in our approach, given the ansatz ${\cal F}$, we immediately obtain the first $r$ derivative of the on-shell Lagrangian, which allows us to obtain the onshell $L(r)$, and hence solve for the $m(r)$ from \eqref{mpr}. It is also instructive to know that
\be
m'' = -\ft12 r L\,,\qquad \hbox{and}\quad
\psi= \frac{2}{q}(m'' r- 2m') \quad \rightarrow \quad a(r) = \frac{2}{q}(m' r- 3 m) + c\,.
\ee
Here $c$ is an integration constant and a pure gauge.

\subsection{Including a cosmological constant}

In this paper, we focus on the black holes that are asymptotic to the Minkowski spacetime. This implies that the scalar potential in the Lagrangian \eqref{escalarvector} has the property $V(\phi=1)=0$. It is straightforward to include a cosmological constant $\Lambda$, by sending
\be
V\rightarrow V + 2\Lambda\,.
\ee
As we pointed it out already, the cosmological constant is simply an integration constant of \eqref{Vprimephi}. The solutions to the scalar and Maxwell fields are unchanged while the metric function $f$ acquires a characteristic (A)dS term:
\be
f\quad\rightarrow\quad  - \ft13 \Lambda\, r^2 + f\,.
\ee

\section{An imperfect resolution of the RN black hole}

In the previous sections, we have constructed several classes of electrically-charged regular black holes that are asymptotic to the RN black hole. This leads to an immediate question: can the RN black hole itself be resolved directly to become regular? As we can see from \eqref{cFsmallr}, a perfect resolution is not possible. This is because for such a regular black hole, ${\cal F}$ must vanish at both asymptotic infinity and at $r=0$.  This implies that ${\cal F}$ must have an extremum at a certain critical $r_*$. However, for the RN black hole, ${\cal F}={\cal F}_0$ and
\be
\fft{d{\cal F}_0}{dr}=\fft{2q^2}{r^5} \ne 0\,,
\ee
for any finite $r_*$. Nevertheless, we can still have an imperfect resolution such that ${\cal F}$ is continuous at $r_*$, but not its derivative. Furthermore, all Riemann curvature invariants of the resolved RN black hole are smooth from $r=0$ to asymptotic infinity.

\subsection{Resolution by NLED}

\begin{figure}[htp]
\centering
\includegraphics[width=140pt]{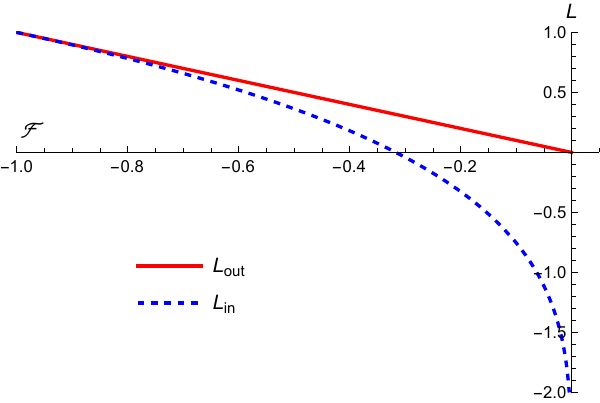}\ \
\includegraphics[width=140pt]{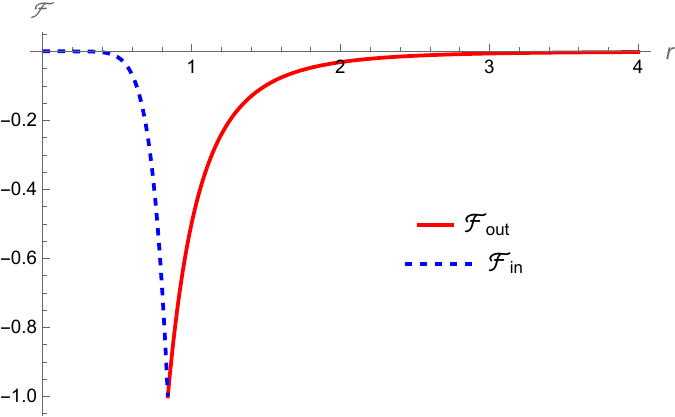}\ \
\includegraphics[width=140pt]{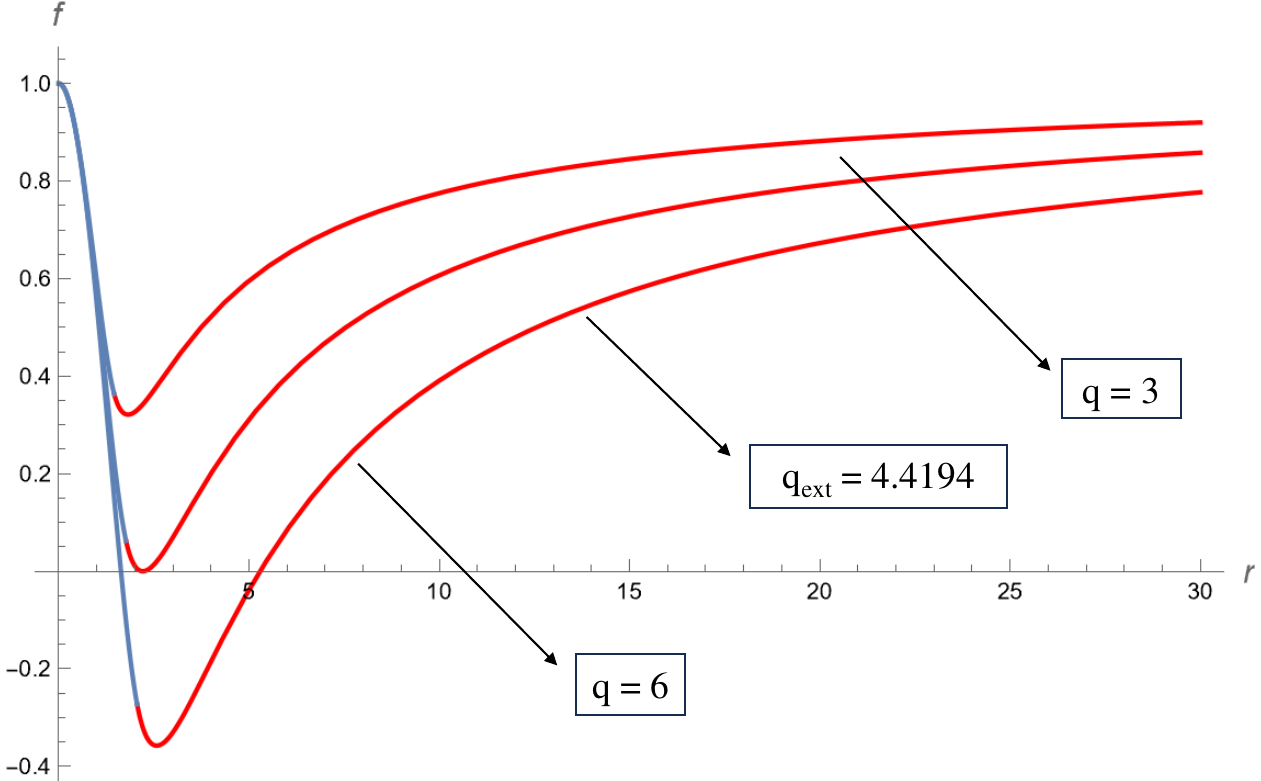}
\caption{\small In these three panels, the red lines show the quantities of the outer region, given by the corresponding quantities of the RN black holes. The blue lines are those of the resolved theory.}\label{figures}
\end{figure}

We first use the NLED $L({\cal F})$ perspective to construct the resolution. In this case, as we have explained earlier, there must be at least two branches of the Lagrangian for the inner and outer regions; therefore, the constructions is really at the level of equations of motion. By construction, the outer region is described by the Maxwell theory, and the corresponding solution is given by
\be
L_{\rm out}=-{\cal F}\,, \quad \rightarrow\quad
{\cal F}={\cal F}_0\,,\qquad f_{\rm out}=1 - \fft{2M}{r} + \fft{q^2}{4r^2}\,.\label{lagout}
\ee
We propose that the Lagrangian for the inner region is
\be
L_{\rm in}= \Lambda_0 + \beta (-\alpha {\cal F})^{\fft14}\,,
\ee
where constant parameters $(\Lambda_0, \alpha, \beta)$ are to be determined. We now require that the two Lagrangians join smoothly at $(\alpha {\cal F}_*)=-1$, namely
\be
L_{\rm out}({\cal F}_*)=L_{\rm in}({\cal F}_*)\,,\qquad
L_{\rm out}'({\cal F}_*)=L_{\rm in}'({\cal F}_*)\,.
\ee
We can determine the couplings $(\Lambda_0, \beta)$, leading to the inner Lagrangian
\be
L_{\rm in}= -\fft{3}{\alpha}  + \fft{4}{\alpha} (-\alpha {\cal F})^{\fft14}\,.
\label{lagin}
\ee
The corresponding solution is
\be
{\cal F}=-\fft{4r^8}{\alpha^3 q^4}\,,
\ee
which corresponds to taking $\nu=3$ in \eqref{cFsmallr}. In the left panel of Fig.~\ref{figures}, we plot the outer and inner Lagrangians as functions of ${\cal F}$. They join at $\alpha{\cal F}=-1$ continuously, but non-smoothly; in fact, the two lines form a cusp at $\alpha{\cal F}=-1$. This is in fact a generic feature of NLED $L({\cal F})$, first observed in \cite{Bronnikov:2000vy}. Our two examples discussed in section \ref{sec:emduality} have the same feature.

The condition $(\alpha {\cal F}_*)=-1$ then implies that
\be
r_* = (\ft12\alpha q^2)^{\fft14}\,.
\ee
The inner regular solution can be easily obtained and it is given by
\be
f_{\rm in}=1 - \fft{r^2}{2\alpha} + \fft{\sqrt2 r^4}{5\alpha ^{\ft32} q}\,.
\ee
Requiring that $f_{\rm out} (r^*) = f_{\rm in}(r^*)$, we find that the critical mass for the resolve RN black hole is
\be
M_{\rm cr}= \fft{2^{1/4} q^{3/2}}{5 \alpha ^{1/4}}\,.\label{RNMcr}
\ee
Note that this mass can also be independently calculated from our previous formula \eqref{MM}, namely
\bea
M_{\rm cr} &=& \int_{0}^q \frac{1}{4} dq \Big(\int_0^\infty \sqrt{-2\cal F} dr\Big) \nn\\
&=& \int_{0}^q \fft14 dq\, \Big( \int_0^{r_*}  \sqrt{-2 {\cal F}_{\rm in}} +\int_{r_*}^\infty \sqrt{-2 {\cal F}_{\rm out}} dr\Big) \,.
\eea
The fact that these two methods yield the same $M_{\rm cr}$ indicates that our resolution of the RN black hole is sound and satisfying the first law. In the middle panel of Fig.~\ref{figures}, we plot the solutions ${\cal F}_{\rm in}$ and  ${\cal F}_{\rm out}$ as functions of the radial coordinate $r$. By construction, they are continuous at $r=r_*$, but not their derivatives. By contrast, note that in the two examples of NLED discussed in section \ref{sec:emduality}, ${\cal F}$ is smooth and infinitely differentiable from $r=0$ to the asymptotic infinity.

On the other hand, for the metric function $f$, not only do we have $f(r_*)_{\rm in} = f(r_*)_{\rm out}$, but also
\be
f'(r_*)_{\rm in} = f'(r_*)_{\rm out}\,,\qquad f''(r_*)_{\rm in} = f''(r_*)_{\rm out}\,,
\ee
indicating that all Riemann tensor polynomial invariants are continuous at $r=r_*$. However, the third and higher derivatives are not continuous. In the right panel of Fig.~\ref{figures}, we give $f$ as function of $r$ for three values: $q=3$, $q=6$ and $q=q_{\rm ext}$, defined by
\be
q_{\rm ext} = \frac{25 \sqrt{\alpha}}{4 \sqrt{2}}\,.
\ee
As we can see from the figure, the $q<q_{\rm ext}$ solutions are regular horizonless spacetimes. The $q\ge q_{\rm ext}$ solutions give regular black holes with generally two horizons, and the saturation corresponds to an extremal black hole with the inner and outer horizons coalescing. For $M=M_{\rm cr}$, given by \eqref{RNMcr}, the original RN metric would have a naked singularity for $q>q_{\rm ext}$, or black holes that are singular at $r=0$ for $q\le q_{\rm ext}$.

It is worth pointing out that, the for the regular resolved RN black hole, the effect of the resolution occurs inside the horizon; the metric, as depicted by the red lines in the right panel of Fig.~\ref{figures}, and the Maxwell field remain unmodified as the usual RN black hole.

\subsection{Resolution by the Maxwell-scalar theory}

The Maxwell-scalar theory \eqref{escalarvector} becomes degenerate if we want to describe the Maxwell theory itself, corresponding to $V=0$ and $\phi=1$. The inner region of the NLED Lagrangian \eqref{lagin} can be equivalently described as
\be
L=-\phi^{-1} {\cal F} + \fft{1}{\alpha} \phi^{\fft13}\,.
\ee
Using the technique discribed earlier, it is easy to verify that we have
\be
\phi = \fft{2\sqrt2}{\alpha^{\fft32} q^3} r^3 = \sqrt{\fft{{\cal F}}{{\cal F}_0}}\,.
\ee
We see that at $r=r_*$, we have $\phi=1$, which connects the inner region to the outer RN metric region. It is easy to verify that the metric function $f$ is identical to that described in the previous subsection.

\section{Conclusions}

In this paper, we considered Einstein gravity minimally coupled to a Maxwell-scalar theory in the form of \eqref{mslag}. For a suitable scalar potential $V(\phi)$, we are able to construct electrically-charged black holes that are geodesically complete. At the level of equations of motion, the Maxwell-scalar theory is equivalent to some NLED; however, it has an advantage of circumventing the no-go theorem of a regular electric black hole under any given $L({\cal F})$ since such a black hole requires multiple branches of $L({\cal F})$.

In the asymptotic weak field region, we would like to insist that the effective NLED can be viewed as an analytic perturbation of the Maxwell theory, in the form of \eqref{wflimit}, which many works in literature do not satisfy. We constructed the EMS theories following the three requirements spelled out in Introduction. We followed two strategies. One is applying the electromagnetic duality to transform the previously known magnetic solutions to electric ones. Our EMS formalism has made the electromagnetic duality simple.

The other approach is to construct the theory directly by the observation that both the theory, namely the scalar potential, and the metric solution are solely determined by an ansatz of ${\cal F}$. Thus by selecting suitable ${\cal F}({\cal F}_0)$, where ${\cal F}_0$ is the value for the RN black hole, we can obtain the desired EMS that admits regular electric black holes. Several classes of the EMS's and their black hole solutions were explicitly given. Our focus was on the asymptotically flat solutions, but we also illustrated explicitly how trivially a cosmological constant could be introduced.

We also analyzed the thermodynamics in the general setting of NLED's and showed that the mass of the regular black holes can be determined solely by the electric potential, without having to know the metric function. Although we have focused on $D=4$ dimensions, our EMS theories are valid in general dimensions. In the appendix, we show that they all admit regular electric black holes in general $D$ dimensions.

The electric RN or EBI black holes have much wider applications than their magnetic counterparts in both astrophysics and the AdS/CFT correspondence. We expect that our construction of the EMS theories and their regular electric black holes can also find applications in those areas.

\section*{Acknowlegement}

This work is supported in part by the National Natural Science Foundation of China (NSFC) grants No.~11935009 and No.~12375052.

\appendix
\section{Regular electric black holes in general dimensions}

In the main text, we focus our attentions on four-dimensional spacetime. In this appendix, we generalize our formalism and solutions to arbitrary dimensions. The EMS theory takes the same form, independent of dimensions:
\bea
{\cal L} = \sqrt{-g} \Big( R - \phi^{-1} \mathcal{F} - V(\phi)\Big)\,,\qquad \hbox{with} \qquad \mathcal{F} = \fft{1}{4} F^{\mu \nu}F_{\mu\nu}\,.
\eea
The equations of motion associated with variations of $(\phi, A_\mu, g_{\mu\nu})$ are
\bea
\delta \phi:&& \quad V'(\phi)=\phi^{-2} \mathcal{F}\,,\qquad\qquad
\delta A_\mu:\qquad \nabla_\mu ( \phi^{-1} F^{\mu\nu})=0\,,\nn\\
\delta g^{\mu\nu}:&&\quad G_{\mu\nu} = \fft{1}{2}\phi^{-1}F^2_{\mu\nu}-\fft{1}{2} \big(\phi^{-1} \mathcal{F} + V(\phi)\big)g_{\mu\nu}\,.
\eea
For spherically-symmetric and static solutions, it can be easily shown that the metric is Schwarzschild-like, i.e., ~$g_{tt}g_{rr}=-1$. We thus consider
\bea
ds^2_D &=& - f(r) dt^2 + \fft{dr^2}{f(r)} + r^2 d\Omega_{D-2}^2\,,\nn\\
\phi &=& \phi(r)\,,\qquad A=a(r) dt\,.
\eea
The Maxwell-equation can be easily solved, giving
\be
\psi\equiv a' = \frac{q\phi}{r^{D-2}}\qquad \rightarrow \qquad \mathcal{F}_{\2} = \phi^2 \mathcal{F}_0\,,
\ee
where $q$ is an integration constant parameterizing the electric charge, and we also define
\be
{\mathcal{F}_0}= -\frac{q^2}{2 r^{2(D-2)}}\,.
\ee
For the general class electric field ansatz \eqref{nnuclass}, the corresponding potential is still given by \eqref{nnuV}, independent of the dimensions.

The dimensional-dependent parts are the explicit solution of ${\cal F}_0$ and the metric profile function $f$, which satisfies
\be
\frac12 (D-2)\frac{f'}{r}+\frac12 (D-2)(D-3)\frac{f-1}{r^2}=-\fft12 V-\frac{q^2 \phi}{4 r^4}\,.
\ee
For the EMS theory with the scalar potential \eqref{nnuV}, we find that the metric function is
\bea
f&=&1-\frac{16\pi M}{(D-2)\Omega_{D-2}\, r^{D-3}}-\frac{ q^2 r^{6-2D}}{ (D-3) (D-2) (D-1)}    \Bigg(\frac{D-3}{2 } \, _2F_1\Big(\frac{1}{n},\frac{\nu}{2};1+\frac{1}{n};-(-\alpha {\cal F}_0)^n\Big)\nn\\
&&-  (D-2) \, _2F_1 \Big(\frac{3-D}{2n (2-D)},\frac{\nu}{2};1+\frac{3-D}{2n (2-D)};-(-\alpha {\cal F}_0)^n\Big) \Bigg).
\eea
The critical mass for the regular black hole can be calculated directly from $f$. It can also be calculated, as we explained in the main text,  directly from ${\cal F}$, without even having to  know $f$:
\bea
M_{\rm cr} &=&  \fft{\Omega_{D-2}}{16\pi} \int_{0}^q  dq \Big(\int_{0}^\infty \sqrt{-2{\cal F}} dr \Big)\nn\\
&=& \frac{(D-2) \Omega_{D-2}\, q^{\frac{D-1}{D-2}} \Gamma \left(1-\frac{D-3}{2 n(2- D) }\right) \Gamma \left(\frac{1}{2} \left(\nu-\frac{D-3}{(D-2) n}\right)\right)}{2^{\frac{7D-13}{2 (D-2)}}  (D-3) (D-1) \pi  \Gamma \left(\frac{\nu }{2} \right)\,\alpha^{\frac{D-3}{2 (D-2)}}}.
\eea
Taking $D=4$ recovers the result in the main text. To include a cosmological constant $\Lambda$ in the Lagrangian, we only need add an appropriate $r^2$ term in $f$, namely
\be
f\rightarrow f - \fft{2}{(D-1)(D-2)} \Lambda\, r^2\,.
\ee
For the general class electric field ansatz \eqref{eclass}, the corresponding potential is still given by \eqref{ve}. The metric function is
\bea
f&=&1-\frac{16\pi M}{(D-2)\Omega_{D-2}\, r^{D-3}}-\nn \frac{2^{\fft1n} r^{D-2}}{\alpha  (D-2) (D-1) n }\nn\\
&& \times\left(2^{\frac{-1}{D-2}}(-2\alpha \mathcal{F}_0)^{\frac{D-1}{2(D-2)}}\Gamma (\frac{D-3}{2n(D-2)},\frac{1}{2} (-\alpha\mathcal{F}_0)^n )+\Gamma (\frac{1}{n},\frac{1}{2} (-\alpha\mathcal{F}_0 )^n)- \Gamma (\frac{1}{n})\right).
\eea
The critical mass for regularity is
\bea
M_{\rm cr} &=&  \fft{\Omega_{D-2}}{16\pi} \int_{0}^q  dq \Big(\int_{0}^\infty \sqrt{-2{\cal F}} dr \Big)\nn\\
&=&\frac{\Omega_{D-2}\alpha^{-\frac{D-3}{2 (D-2)}} 2^{\frac{D (-n)+D+n-3}{2 (D-2) n}} q^{\frac{D-1}{D-2}} \Gamma \left(\frac{3-D}{4 n-2 D n}\right)}{16 \pi(D-1) n}.
\eea

\end{document}